\documentclass[sigconf]{acmart}
\AtBeginDocument{%
  }

\setcopyright{acmlicensed}
\copyrightyear{2025}
\acmYear{2025}
\acmDOI{XXXXXXX.XXXXXXX}

\acmConference[Augmented Humans 2025]{Augmented Humans 2025}{March 16--20,
  2025}{Abu Dhabi, UAE}
\acmISBN{978-1-4503-XXXX-X/18/06}

\usepackage{multirow}
\newcommand{\commentout}[1]{}

\usepackage{enumitem}
\setlist[description]{leftmargin=\parindent,labelindent=0pt}
\setlist[itemize]{leftmargin=\parindent,labelindent=0pt}
\usepackage{ulem} 

\newcommand{\add}[1]{#1}
\newcommand{\delete}[1]{}

\begin{document}


\author{Jun Rekimoto}
\orcid{0000-0002-3629-2514}
\affiliation{%
 \institution{The University of Tokyo}
\city{Bunkyo-ku}
\state{Tokyo}
\country{Japan}
\postcode{113-0033}
}
\affiliation{%
\institution{Sony CSL - Kyoto}
\city{Kyoto-shi}
\state{Kyoto}
\country{Japan}
}
\email{rekimoto@acm.org}

\renewcommand{\shortauthors}{J.Rekimoto}

\title{GazeLLM: Multimodal LLMs incorporating Human Visual Attention}

\begin{abstract}

Large Language Models (LLMs) are advancing into Multimodal LLMs (MLLMs), capable of processing image, audio, and video as well as text. Combining first-person video, MLLMs show promising potential for understanding human activities through video and audio, enabling many human-computer interaction and human-augmentation applications such as human activity support, real-world agents, and skill transfer to robots or other individuals. However, handling high-resolution, long-duration videos generates large latent representations, leading to substantial memory and processing demands, limiting the length and resolution MLLMs can manage. Reducing video resolution can lower memory usage but often compromises comprehension. This paper introduces a method that optimizes first-person video analysis by integrating eye-tracking data, and proposes a method that decomposes first-person vision video into sub areas for regions of gaze focus. 
By processing these selectively gazed-focused inputs, our approach achieves task comprehension equivalent to or even better than processing the entire image at full resolution, but with significantly reduced video data input (reduce the number of pixels to one-tenth), offering an efficient solution for using MLLMs to interpret and utilize human skills.
\commentout{
prioritizing high resolution for gaze-focused areas and lower resolution for peripheral regions. We demonstrate that in domains such as cooking and traditional performing arts, this approach preserves comprehension quality while reducing memory consumption, offering an efficient solution for using MLLMs to interpret and utilize human skills.
}
\end{abstract}





\begin{teaserfigure}
\centering
 \includegraphics[width=.85\textwidth]{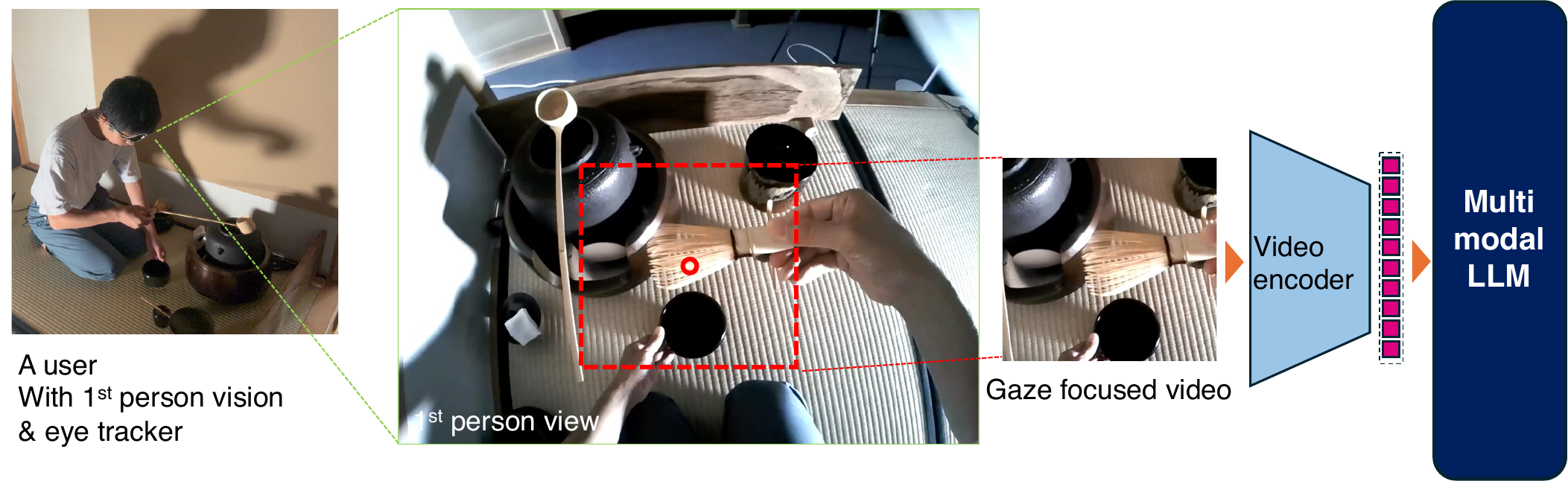}
  \caption{GazeLLM achieves task descriptions of equal or superior quality to full-sized videos, despite using only one-tenth of the pixels around the user's gaze point. }
  \label{fig:teaser}
\end{teaserfigure}


\maketitle

\section{Introduction}


Large Language Models (LLMs) have become foundational in natural language processing, underpinning many advances in understanding and generating human language. Recently, these capabilities have expanded to encompass multimodal data, leading to the development of Multimodal LLMs (MLLMs). These models process not only text but also images and time-sequential information, such as video and audio~\cite{yin2024surveymultimodallargelanguage}. MLLMs represent a substantial leap toward Artificial General Intelligence (AGI) as they integrate multisensory perception—relying on vision, sound, and other sensory modalities—mirroring the complex ways humans process information.

From a Human Computer Interaction (HCI) and Human Augmentation (HA) perspective, MLLMs also offer various opportunities. If such models can recognize the world in ways similar to humans, a range of applications becomes possible. These include technologies that can record and understand skilled human actions for transfer to others, assess skill development, recognize real-world behaviors to provide personalized assistance and assist individuals with disabilities by augmenting their sensory perception of the environment.
In particular, by enabling MLLMs to interpret video data obtained from first-person vision, it becomes possible to understand and explain the actions and skills of humans performing specific tasks.

However, the computational burden of MLLMs remains a significant challenge. The transformer models at the core of LLMs require memory proportional to the square of the input length, which increases both processing time and memory demands. This problem is especially acute in handling image or movie data.

The primary image encoder used in MLLMs,
such as Vision Transformers (ViT)~\cite{dosovitskiy2021imageworth16x16words}, images are divided into grid-like regions, with each region assigned a token. This leads to an increase in token count as the number of pixels increases. Reducing the token count by lowering the image resolution or downsampling the frame rate is possible; however, this introduces a trade-off, as the resolution may become insufficient for recognizing the environment adequately. For instance, tasks such as reading text on a sign directly in front of the user require image resolution sufficient for OCR processing.

On the other hand, human visual perception does not process all visual information equally. Through eye movements, humans selectively focus on specific regions while information in peripheral vision is perceived at a lower resolution, or only reacts to motion. This visual attention mechanism allows humans to balance the precision of world understanding with the efficiency of processing visual input.

Our study leverages this natural visual attention mechanism to reduce the processing load of dynamic visual information in MLLMs.

Here, we focus on 1st-person video as an input to MLLM, to recognize the content of the task from it. In addition, using an eye-tracking equipped first-person vision camera,  we propose a mechanism that decomposes first-person vision video into high-resolution areas for regions of gaze focus. 

By processing these selectively focused inputs, we demonstrate that our approach achieves comprehension equivalent to or even better than processing the entire image at full resolution, but with significantly reduced data input.

The contributions of this paper are as follows:

\begin{enumerate}
\item We confirmed that applying 1st-person vision to MLLM enables accurate comprehension of task procedures by examining six types of real-world activities including cooking, repair, first aid, and sports.
\item User evaluations demonstrated that utilizing partial video cropped around the gaze point from 1st-person vision (which we call `GazeLLM') achieves task descriptions of equal or superior quality to full-sized videos, despite using only one-tenth of the pixels.
\item Additional quantitative evaluations (BLEU, ROUGE, Sentence-BERT, and LLM-based methods) further confirmed that gaze-centered cropped videos effectively support task description generation.
\end{enumerate}

These findings suggest that GazeLLM (a gaze-based LLM) is a promising framework for real-world assistance systems integrating AI and wearable technology.

\section{Related Work}


\subsection*{Multimodal Large Language Models (MLLMs)}

The field of Multimodal Large Language Models (MLLMs) is an area of LLM research that has seen rapid advancements in recent years~\cite{yin2023survey}. In addition to static images and audio, some MLLMs can now process video content~\cite{fu2024video,fu2024vita}. Notably, Gemini 1.5 Pro~\cite{geminiteam2024gemini15unlockingmultimodal} is capable of handling over an hour of video input, while Qwen2-VL~\cite{Qwen2VL} can operate as a locally run MLLM system.

Currently, these MLLMs provide dedicated encoders for video processing, frequently utilizing Vision Transformer (ViT)~\cite{dosovitskiy2021imageworth16x16words} or its variations. In the ViT approach, images are divided into a grid of two-dimensional patches, each assigned a latent representation vector. This vector sequence can be treated as a one-dimensional token series, enabling the integration of visual inputs into the LLM alongside other modalities. However, this method introduces a trade-off: as pixel count increases, so does the number of tokens, which may either increase computational load or require reducing image resolution, potentially compromising the accuracy of visual comprehension.

\begin{figure*}
\centering
  \includegraphics[width=.9\textwidth]{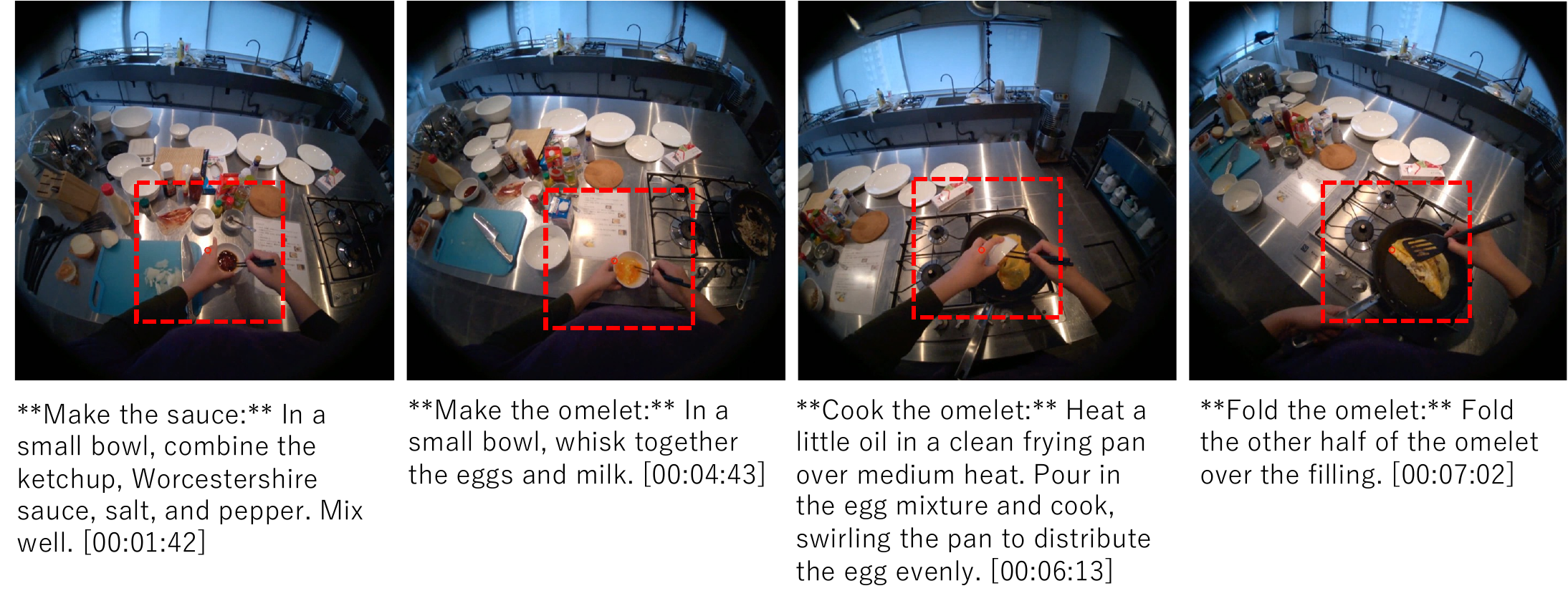}
  \caption{Video task description using GazeLLM: Only the area cropped as a partial rectangle centered on the gaze point in the 1st-person video (indicated by the red dashed outline) is used as input to the MLLM for task verbalization. The output from the LLM includes the task's description and the corresponding video timestamp. The generated description is displayed alongside the video footage at the indicated time.}
  \label{fig:om}
\end{figure*}

\subsection*{Activity Recognition Using Gaze}

Following Bolt’s pioneering work~\cite{10.1145/800049.801811}, numerous studies have explored the use of gaze information in user interfaces. GEAR, for instance, utilizes gaze data to infer the task category for AR-headset users, employing machine learning methods such as SVM and Random Forest~\cite{10.1145/3025453.3025821}.

\add{
In recent years, there have been efforts to combine gaze information with neural networks. This research includes studies on the similarity between human attentional mechanisms and those employed in neural networks~\cite{wang2024probinglargelanguagemodels,mazzamuto2024gazingmisstepsleveragingeyegaze}, investigations aimed at estimating human attention through neural networks~\cite{lin2024multimodallargelanguagemodels}, and attempts to enhance the inference accuracy of neural networks by leveraging human attention~\cite{sugano2016seeing,sood23_gaze,konrad2024gazegptaugmentinghumancapabilities,10.1145/3659623,10.1145/3025453.3025821,Ni_2023}.
}
The Multimodal Human-like Attention Network (MULAN)~\cite{sood23_gaze} adopts human-like visual attention based on image saliency in Visual Question Answering (VQA). GazeGPT~\cite{konrad2024gazegptaugmentinghumancapabilities} improves VQA accuracy by focusing on regions where the gaze is concentrated, using static images and gaze data as input. 
\add{G-VOILA~\cite{10.1145/3659623} is also an interactive system that combines speech dialogue and visual and gaze information to augment conversational context based on the area observed by the user. For example, when a user views an ingredient and inquires about its preparation method, the system provides a recipe suitable for that ingredient.
Ni et al. demonstrated gaze informatin can be used to predict human behavior~\cite{Ni_2023}. 
}

In our study, we incorporate gaze information not only from static images but also from video, to understand and utilize demonstrations of human skills through video and gaze data, in conjunction with multimodal large language models (MLLMs).

Wang et al. investigate the similarities between human gaze patterns and the attention mechanism in LLMs~\cite{wang2024probinglargelanguagemodels}. It has identified parallels between human eye movements when interpreting text and the attention mechanism within LLMs. 
\add{
Mazzamuto et al. demonstrates unsupervised human misstep detection by analyzing differences between real gaze and estimated gaze~\cite{mazzamuto2024gazingmisstepsleveragingeyegaze}.  
}
While our study primarily utilizes, rather than predicts, gaze information from users to improve processing efficiency, future research may explore the similarities between human gaze mechanisms and the attention mechanism in multimodal-LLMs.


EgoScanning employs cues such as gestures and conversations to adjust the density of timelines for efficient browsing of first-person videos~\cite{10.1145/3025453.3025821}. While this technology enhances human viewing efficiency, our study explores its utility as an aid for AI in interpreting long-duration videos.

\subsection*{Activity Dataset with First-person Videos and Gaze}

The Ego-Exo4D project aims to build a large-scale dataset for first-person vision~\cite{Grauman_2024_CVPR}. 
Multiple research institutions are collaborating to collect data across various task domains. The project's primary objective is to establish and provide a data infrastructure for studying human behavior. However, specific attempts that utilize multimodal-LLMs in combination with gaze, as proposed in this study, have not been reported.
\add{
VQA-MHUG~\cite{sood-etal-2021-vqa} is also a dataset to study multimodal neural attention in Visual Question Answering (VQA). 
It consists of 49-participant dataset to analyze similarity between human attention and multi-modal neural attention.}

\add{
\subsection{Devices for Acquiring First-Person video and gaze Information}

In practical work tasks, the ease with which devices capturing first-person perspectives and visual information can be worn is paramount. Meta has developed its proprietary glasses-type `Aria' device for constructing the Ego-Exo4D dataset~\cite{engel2023projectarianewtool}. In our experiments, we employed the Pupil labs `NEON', a device capable of recording first-person video, gaze data, and audio, and which is also compatible with corrective lenses~\cite{pupillabNEON}. Because gaze measurements can be performed without requiring calibration, the initial operational burden is substantially reduced. 
Moreover, certain VR/AR headsets, such as Apple VisionPro, employ gaze input, indicating the potential for supporting real-world tasks through first-person vision, gaze tracking, and video pass-through functionalities~\cite{10.1016/j.cag.2024.103909}.
Consequently, we posit that work support using smart glasses equipped with first-person perspective and visual tracking capabilities will become widely available in the near future.
}

\section{Examples of Gaze-enables Wearable Interaction with GazeLLM}

In GazeLLM, the 1st-person video serves as input to enable the MLLM to understand the tasks being performed by the user, facilitating video-based guidance and question answering. Gaze information from the user’s 1st-person view is utilized to crop only the surrounding region, which is then provided as input to the LLM. Compared to using the entire 1st-person video, the input in our experiments was reduced to approximately one-tenth of the original pixel count. Fig~\ref{fig:om} shows video description generated with  MLLM (Gemini1.5pro) by using gaze-focused region.

\begin{figure*}
\centering
  \includegraphics[width=.8\textwidth]{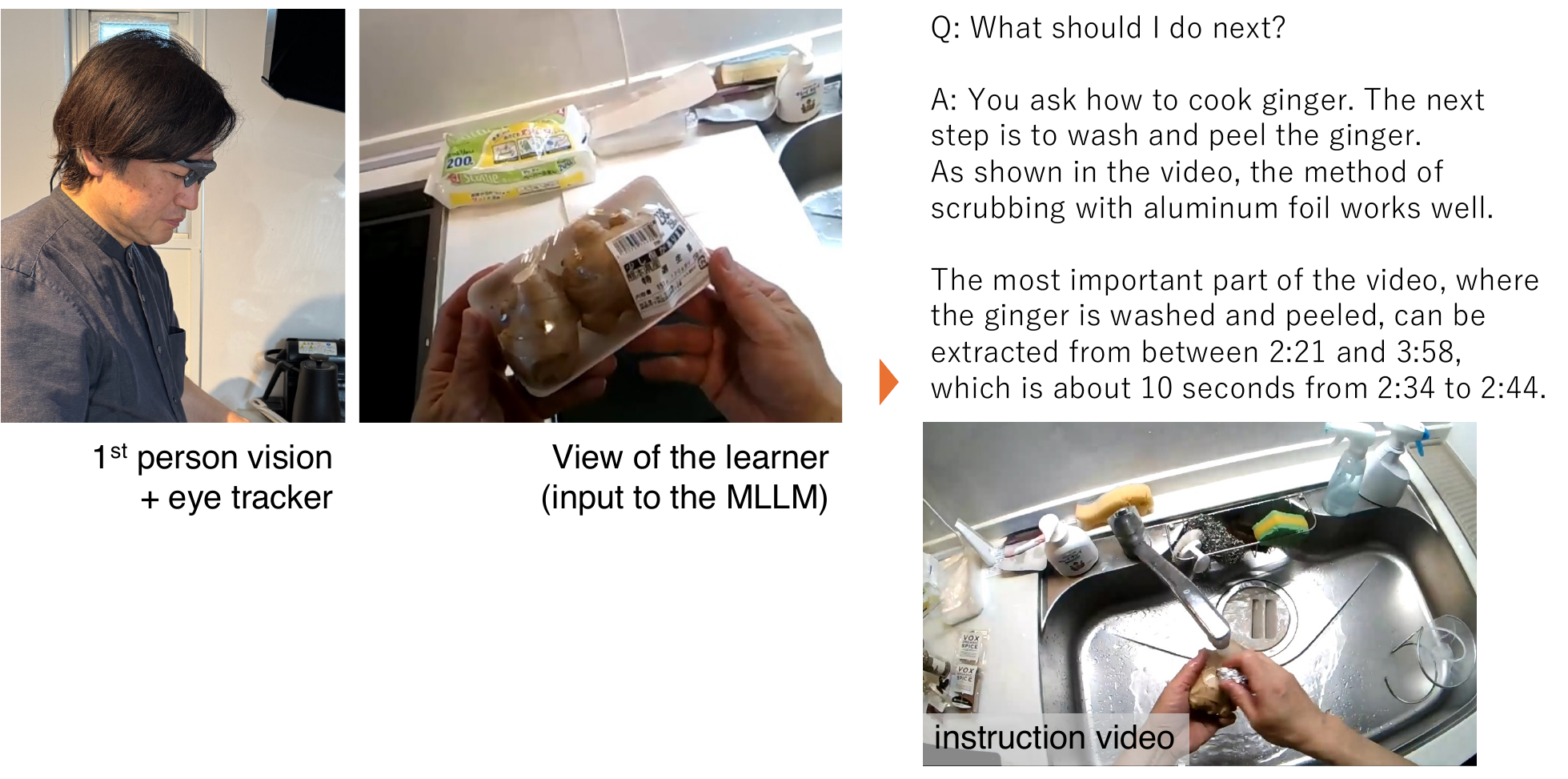}
  \caption{Example of the ``stop-and-ask'' application: A user wearing a 1st-person vision headset pauses during a task and asks the system, ``What should I do next?'' Based on the 1st-person vision video recorded up to that point, the system identifies the next required step and provides guidance. It also references the pre-recorded instructional video and indicates the corresponding playback timestamp.}
  \label{fig:stop}
\end{figure*}

Another example of GazeLLM is called the ``stop and ask'' approach (Fig~\ref{fig:stop}). With this approach, first, an instructor pre-trains the LLM on a sequence of cooking steps using a 1st-person video. A learner then attempts to follow the same type of recipe, equipped with headsets capable of recording 1st-person vision. The learner proceeds partway through the cooking process, pauses, and asks to the system (e.g., ``What should I do next?'' or ``Which utensil I should use now?''). The system successfully identifies the learner’s progress within the sequence of steps provided by the instructor and accurately indicates the next action. Other possible inquiry examples are shown in Table~\ref{tab:inc}.

\begin{table*}
\begin{tabular}{l|l}
\multicolumn{1}{c|}{Inquiry types} & \multicolumn{1}{c}{Typical Inquiries} \\
\hline
Inquiry about operation details 
& 
{\it Please show me the procedure for performing ***.} \\
Inquiry about differences in operations 
& 
{\it Where are the differences between this operation and that one?} \\
Inquiry about skill proficiency & {Please point out any areas that need improvement in this operation.}\\
Inquiry about the next step in the operation &
{\it I've done it up to this point. Please show me what to do next.}\\

Monitoring and advice for the work
&
{\it Let me know if there are any mistakes in the upcoming operations.}\\

&
{\it Is it correct that the next ingredient to add is this one?}\\
\hline
\end{tabular}
\caption{Inquiry types to GazeLLM}
\label{tab:inc}
\end{table*}

\begin{figure*}
\centering
  \includegraphics[width=0.6\textwidth]{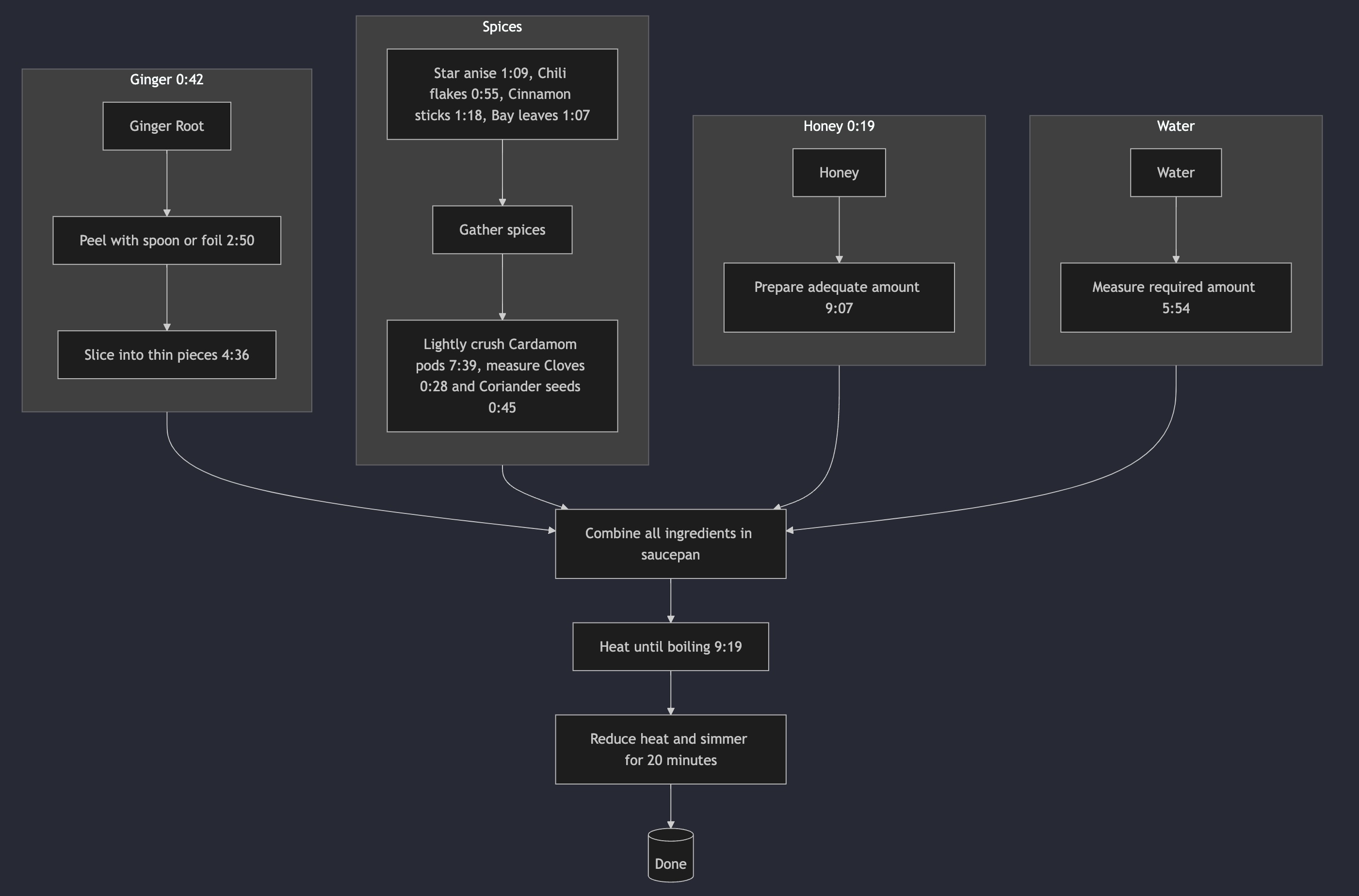}
  \caption{Dataflow generation from cooking demonstration.}
  \label{fig:flow}
\end{figure*}

It is also possible to present tasks in a format other than text. Based on a cooking demonstration with 1st-person video and voice annotation, the MLLM automatically generates the ingredient preparation process as a dataflow diagram (Fig~\ref{fig:flow}). By illustrating branching tasks that can be performed in parallel, the procedure becomes easier to visualize and understand.

These applications demonstrate the potential of combining wearable systems with MLLMs. Users can utilize 1st-person vision to allow the MLLM to recognize their task context (such as cooking, bike repair, or first aid) and receive guidance specific to their current situation.

In this way, the integration of Multimodal LLMs with wearable systems, using 1st-person vision as a contextual input, opens up the possibility of employing AI as an agent in real-world applications.

\section{Evaluation}

To confirm the effectiveness of the GazeLLM's gaze-focus approach,
we evaluated the generated video description texts using both numerical metrics and human-based crowdsourced assessment.

\subsection{Dataset}

For our evaluation, we utilized the Ego-Exo4D dataset~\cite{Grauman_2024_CVPR}. The Ego-Exo4D dataset contains gaze information captured using the `Aria' glasses~\cite{engel2023projectarianewtool} developed by Meta, 1st-person vision images, and third-person perspective footage obtained from external cameras. This dataset encompasses a wide variety of scenarios, including cooking, repair, and sports, making it a valuable resource for research.

\begin{figure*}
\centering
  \includegraphics[width=0.8\textwidth]{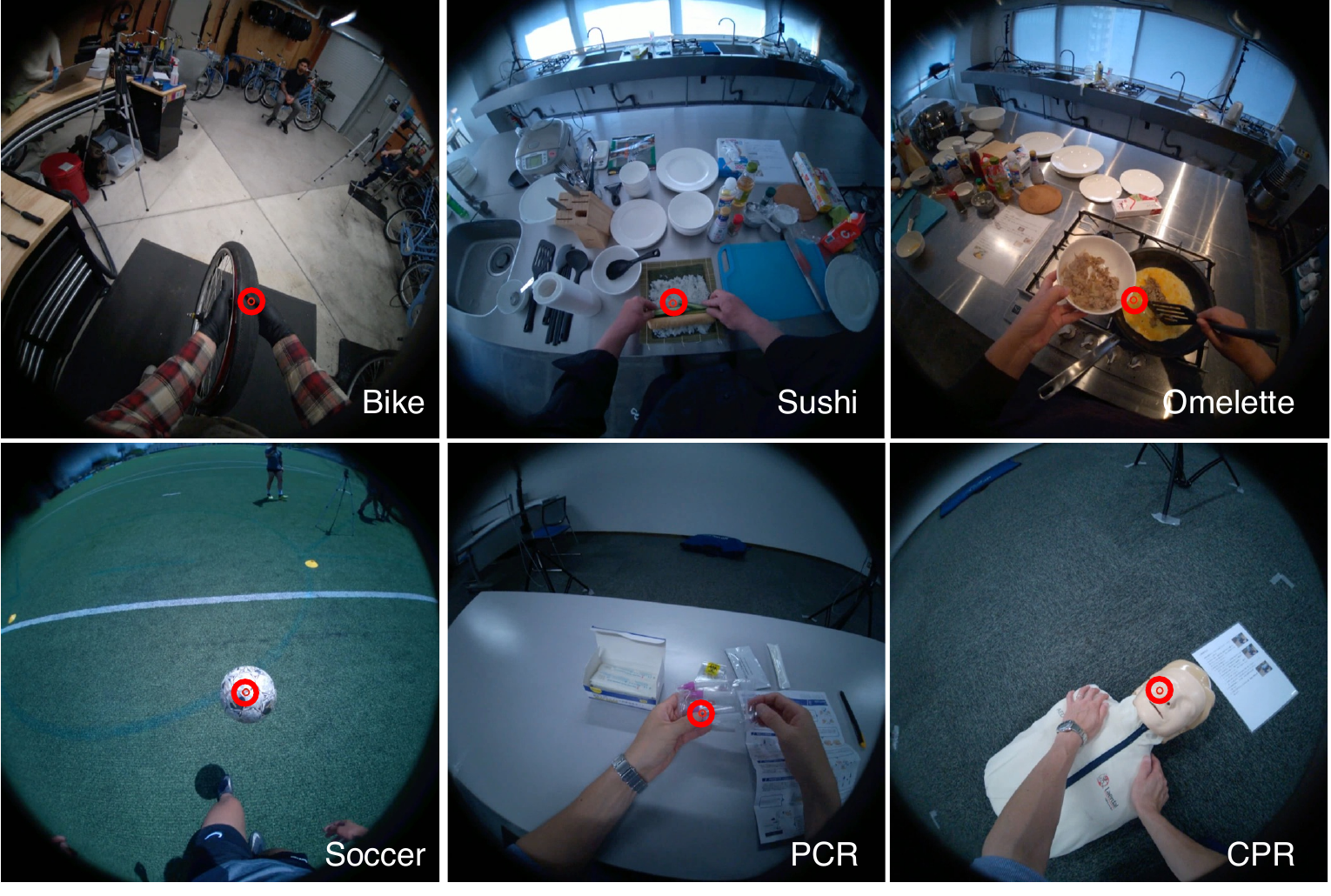}
  \caption{Evaluation tasks: labeled 'Bike', 'Sushi', 'Omelette', 'Soccer', 'PCR' (Polymerase Chain Reaction test), and 'CPR'(cardiopulmonary resuscitation), Red circles indicating viewpoints are attached for reference.}
  \label{fig:tasks}
\end{figure*}

\begin{table*}
    \centering
    \begin{tabular}{c|l|c|r}
label & \multicolumn{1}{c|}{task} & \# of videos & \multicolumn{1}{c}{duration (sec)} \\
\hline
{\bf Bike} & bike repair & 20 & 181.8 ($162.5$) \\
{\bf Sushi} & sushi preparation & 14 & 627.7 ($106.5$) \\
{\bf Omelette} & omelette preparation & 14 & 783.0 ($269.9$) \\
{\bf Soccer} & soccer activities & 19 & 132.6 ($81.0$) \\
{\bf PCR} & Polymerase Chain Reaction (PCR) testing equipment preparation & 53 & 253.8 ($86.5$) \\
{\bf CPR} & cardiopulmonary resuscitation (CPR) training & 17 & 95.4 ($27.1$) \\
\hline
    \end{tabular}
    \caption{Tasks for evaluation: video length distribution}
    \label{tab:tasks}
\end{table*}

In this evaluation, we employed only the 1st-person video and gaze information. To ensure task diversity, we selected six distinct task categories (Fig.~\ref{fig:tasks}). Table~\ref{tab:tasks} presents the data count, video duration, and standard deviation for each task. The number of total tasks was 137.

\commentout{
\begin{description}
\item[Bike] Footage of bike repair
\item[Sushi] Footage of sushi preparation
\item[Omelette] Footage of omelette preparation
\item[Soccer] Footage of soccer activities
\item[PCR] Footage of Polymerase Chain Reaction (PCR) testing preparation
\item[CPR] Footage of cardiopulmonary resuscitation (CPR) training
\end{description}
}

For each task, we prepared three types of data (Fig~\ref{fig:fullgaze}):

\begin{figure*}
\centering
  \includegraphics[width=0.7\textwidth]{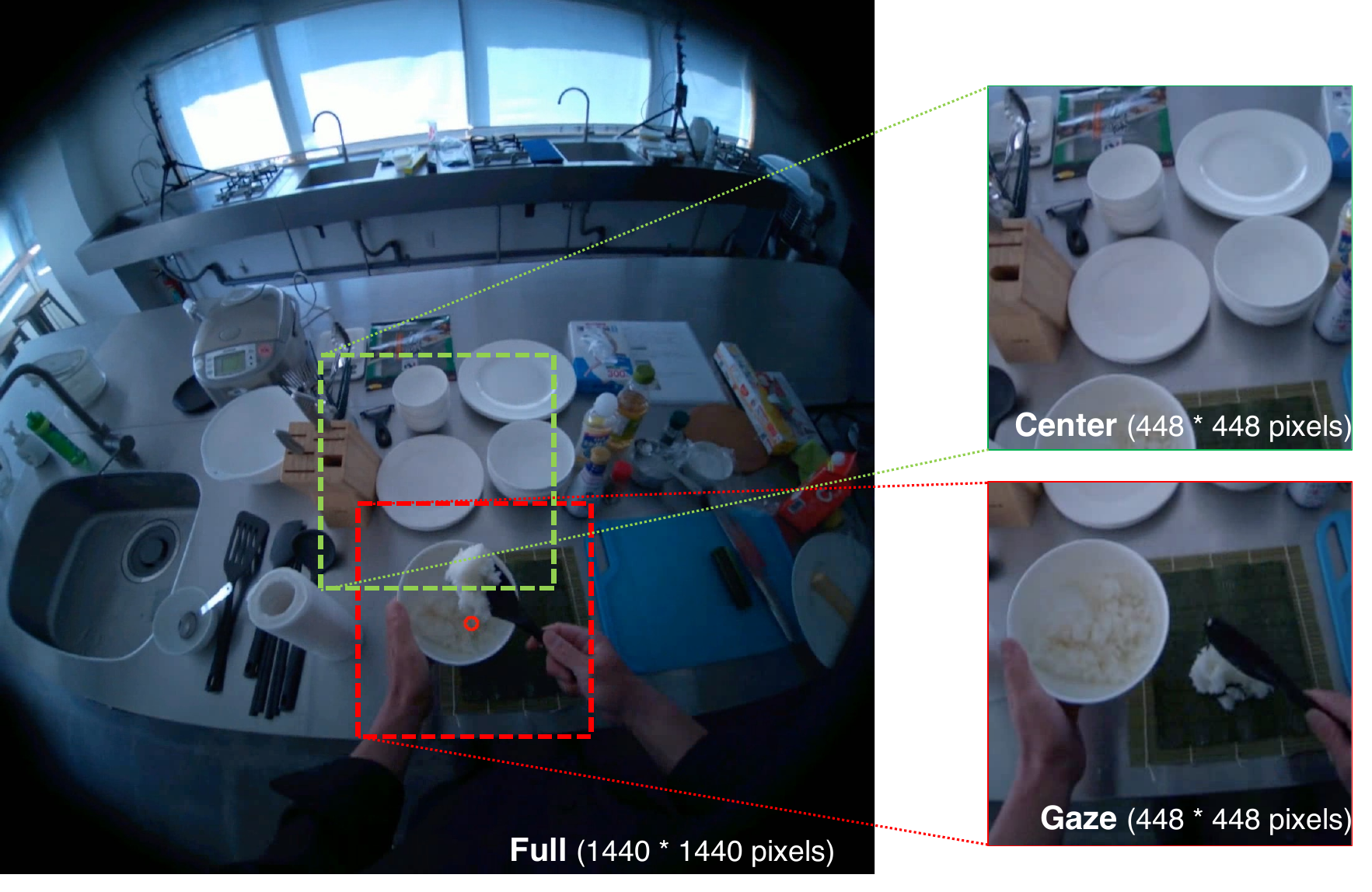}
  \caption{Video types for evaluation (Full: Full field of view, Gaze: Video cropped around the gaze point, Center: Video cropped from the center of the image)}
  \label{fig:fullgaze}
\end{figure*}

\begin{description}
\item[Full] The first-person perspective video provided by Ego-Exo4D, with an image size of $1440 \times 1440$ pixels, is used as-is, but with the video frame rate reduced to 1 fps. No gaze information is included.
\item[Gaze] This data is a cropped version of Full, centered around the gaze point in a $ 448 \times 448$ rectangular area. The frame rate is maintained at 1 fps, as in {\bf Full}.
\item[Center] This data is cropped from the center of the Full image, also in a $448 \times 448$ rectangular area, with a frame rate of 1 fps, as in {\bf Full}.
\end{description}

In order to clearly determine the differences in video extraction, none of the videos include audio. The pixel count of Gaze and Center ($448 \times 448 = 200,704$) represents $10.3\% $ of that of Full ($1440 \times 1440 = 2,073,600$). The evaluation metric is a comparison of the quality of video descriptions generated with these conditions.

For the MLLM under evaluation, we selected Gemini 1.5 Pro~\cite{geminiteam2024gemini15unlockingmultimodal}. Gemini 1.5 Pro is an Multimodal-LLM capable of processing up to approximately one hour of video (with the video frame rate down-sampled to 1 fps). \add{
Using the Gemini API, sessions are explicitly separated to ensure that the results of the previous session do not persist.
}
Descriptions of the videos are generated from the dataset using the prompt below:

\begin{verbatim}
Clear chat history and begin a new session.
Forget all the previous information and only use 
the information from this video. 

Please create a written procedure in English for the work
in this video. I would like to insert an image of the work
in progress in the instructions to make it easier to see,
so please write the time of the frame of the video to be
inserted.
\end{verbatim}

\subsection{Evaluation Results Based on Users}

\begin{figure}
\centering
  \includegraphics[width=0.5\textwidth]{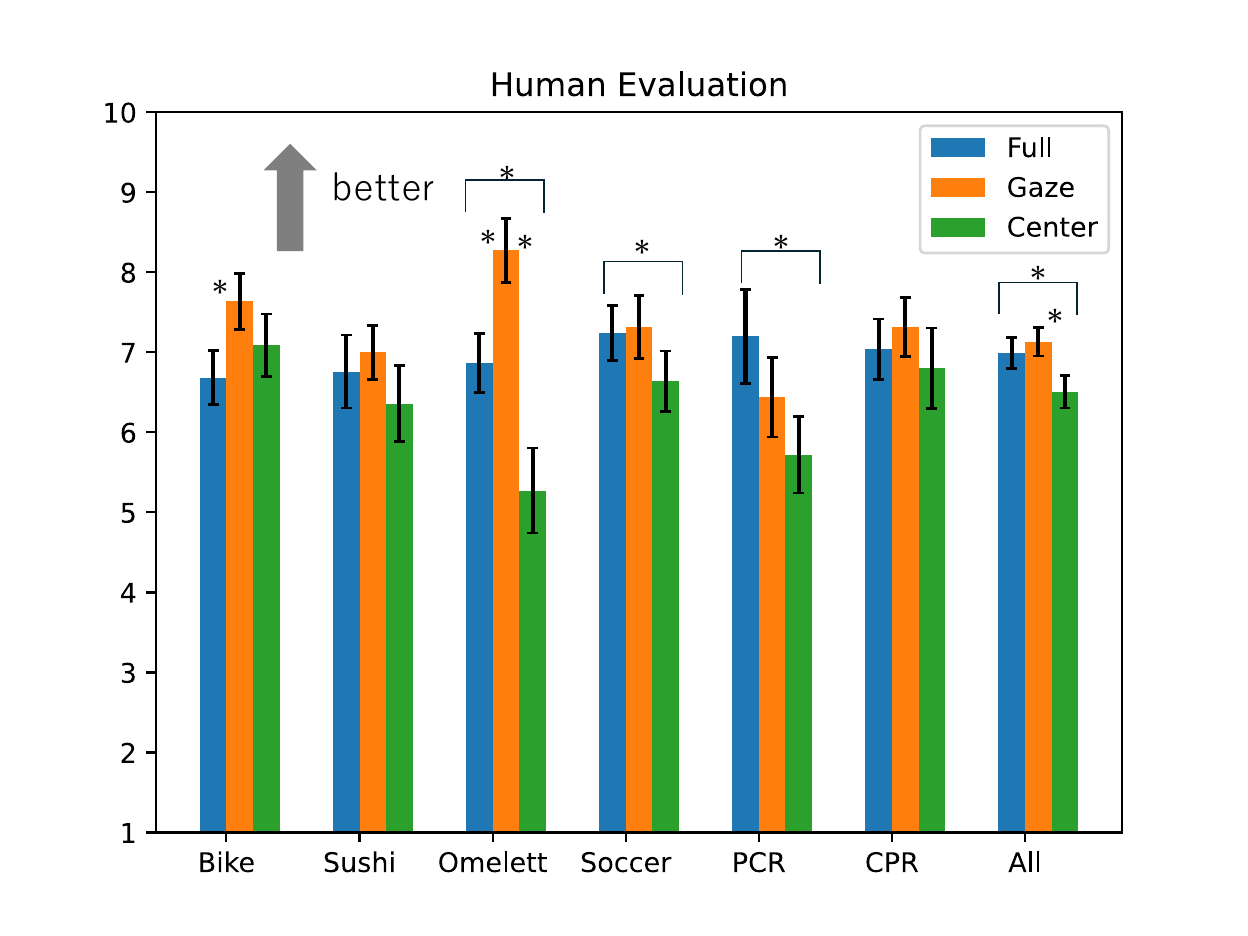}
  \caption{Video description evaluation results by humans\add{(*: $p < 0.05$)}}
  \label{fig:human}
\end{figure}

\begin{figure*}
\centering
  \includegraphics[width=.9\textwidth]{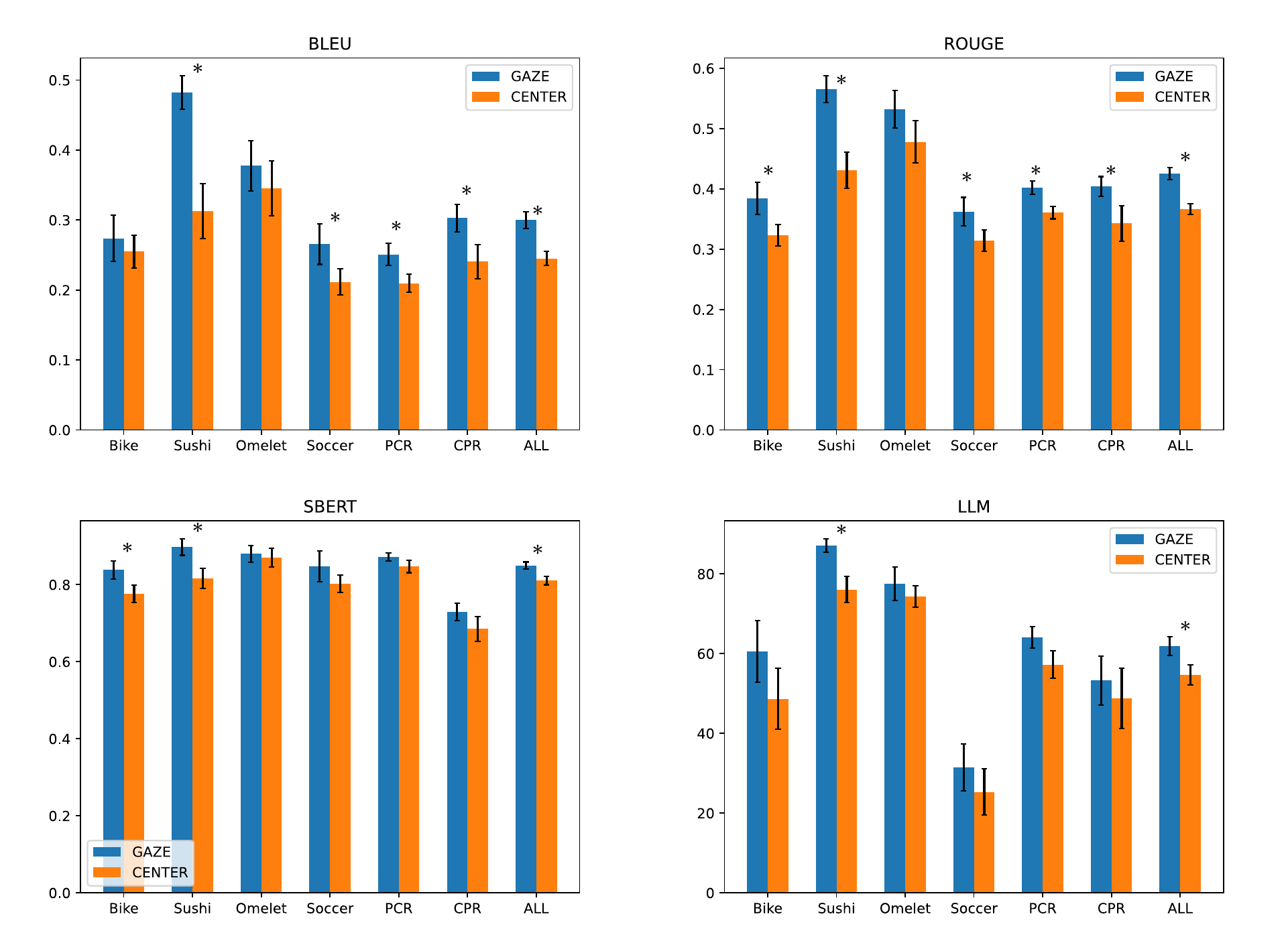}
  \caption{Video description evaluation by BLEU, ROUGE, SBERT (sentence-BERT) and LLM \add{(*: $p < 0.05$)}}
  \label{fig:graph}
\end{figure*}

First, we conducted a human evaluation experiment. Participants rated the descriptions generated by Gemini under the Full, Gaze, and Center conditions, following a viewing of the Full video, on a 10-point scale.

The experiment was conducted using the Prolific crowd-sourcing platform~\cite{prolific}. We recruited 25 online participants, who provided a numerical score on a 10-point scale and free-text feedback on the three types of text descriptions (Full, Gaze, Center) about the full video. To prevent anticipation effects, text description types were not disclosed and the order of video descriptions was randomized. Since the descriptions were in English, participants were recruited based on English proficiency. The average age of participants was 29.5 years ($8.4$).

The results are shown in Fig~\ref{fig:human}. \add{Pair-wise t-test confirmed Full and Gaze are significantly better than Center, and there was no significant difference between Full and Gaze ($p < 0.05$)},  but the Gaze condition received the highest average scores in five out of six types of tasks. Although participants only viewed the Full-condition video, it is inferred that they tended to rate descriptions related to areas of focus (where gaze was fixated) more highly, while giving lower ratings to descriptions of areas irrelevant to the task (related to regions outside the gaze focus). This result strongly suggests Gaze video convey accurate information about the task, and MLLM can recognize task more accurately than the Full and Center video conditions.

\subsection{Evaluation Results Based on Numerical Metrics}

We then conducted the numerical evaluation aimed to assess the accuracy of descriptions generated under the Gaze and Center conditions, compared to those generated under the Full condition. We employed the following four metrics for this purpose:

\begin{description}
\item[BLEU] The BLEU score~\cite{papineni-etal-2002-bleu} is one of the most widely used methods for machine translation evaluation. It operates on the principle that the closer the machine-generated text is to a professional translator's output, the higher its accuracy. In this evaluation, we used BLEU scores to measure the accuracy of descriptions generated under the Gaze and Center conditions, taking the Full condition descriptions as the reference.
\item[ROUGE] ROUGE (Recall-Oriented Understudy for Gisting Evaluation)~\cite{lin-2004-rouge} is also widely used in natural language processing to assess machine-generated summaries or translations by comparing them with human-created references. For this evaluation, we measured the accuracy of the Gaze and Center descriptions against the Full condition descriptions using ROUGE-L, which is based on the Longest Common Subsequence (LCS). ROUGE-L provides an evaluation of sentence-level similarity by counting co-occurring words in order between the target and reference summaries.
\item[SBERT] Sentence-BERT (SBERT)~\cite{reimers2019sentencebertsentenceembeddingsusing} is a version of BERT~\cite{devlin-etal-2019-bert} fine-tuned for sentence comparison, to evaluate semantic similarity. SBERT utilizes Siamese and triplet network structures to derive meaningful sentence embeddings that can be compared using cosine similarity. This approach is particularly well-suited for measuring similarity across multi-sentence descriptions. Here, we assessed the similarity between descriptions generated under the Full condition and those generated under the Gaze and Center conditions. A higher similarity indicates that the generated description aligns closely with the Full condition output.

\item[LLM] We conducted an evaluation using the LLM itself. Specifically, we used ChatGPT-4o~\cite{openai2023gpt4} to score Gaze and Center condition descriptions against the Full condition description on a scale of 0 to 100, based on the following prompt:
\end{description}

\begin{verbatim}
I will provide you with two texts: Text A and Text B. 
Please evaluate how well the content described in Text A
is also covered in Text B.  Assess the similarity between
the two texts based on whether Text B includes the key
information, ideas, and explanations found in Text A,
regardless of wording or phrasing differences.

Please provide a score from 0 to 100, where:
- 100 means Text B completely covers all the key points 
  and information from Text A,
- 0 means Text B does not cover any of the key points
  from Text A,
- A score between 0 and 100 represents partial coverage.

Show score as '** Score:50 **'

After assigning a score, explain the reasoning for
your score in a few sentences.

### Text A:
{text_1}

### Text B:
{text_2}
        \end{verbatim}

The evaluation results are shown in Fig~\ref{fig:graph}. For all evaluation conditions and task items, the scores under the Gaze condition was significantly superior than those of the Center condition \add{(pair-wise t-test, $p < 0.05)$)}. This suggests that gaze-based cropping captures task-relevant content more accurately than simple center-based cropping.

Both BLEU and ROUGE are statistical measures based on n-grams, responding to identical vocabulary and phrases. In contrast, SBERT uses cosine similarity within a sentence embedding space, evaluating sentence similarity at a broader level beyond specific vocabulary. The LLM condition relies on an LLM, which is assumed to provide a comprehensive assessment based on overall sentence comprehension. These four distinct metrics enable a mechanistic comparison between the Full condition and other conditions.

In these evaluations, by definition, the Full condition achieves a perfect score (BLEU, ROUGE, and SBERT: 1.0; LLM: 100 points). However, it is important to note that captions generated under the Full condition do not necessarily represent the most accurate ground truth. Given the wide field-of-view nature of first-person video, descriptions may sometimes include objects in the peripheral view that are not directly related to the task.

\begin{table}
    \centering
    \begin{tabular}{c|r|r|r}
    task & \multicolumn{1}{c|}{Full (std)} & \multicolumn{1}{c|}{Gaze (std)} & \multicolumn{1}{c}{Center (std)} \\
    \hline
{\bf Bike } & 1312.7 ($521.8$) & 1419.8 ($563.4$) & 1561.7 ($432.0$)\\
{\bf Sushi } & 1968.1 ($169.5$) & 1996.3 ($192.3$) & 1711.9 ($543.0$)\\
{\bf Omelet } & 1600.2 ($244.4$) & 1656.6 ($316.5$) & 1468.2 ($191.2$)\\
{\bf Soccer } & 2107.2 ($358.8$) & 1728.1 ($336.0$) & 1768.1 ($364.6$)\\
{\bf PCR } & 1522.5 ($403.9$) & 1386.8 ($481.1$) & 1342.2 ($513.3$)\\
{\bf CPR } & 1845.6 ($292.1$) & 1866.5 ($394.8$) & 1593.7 ($438.8$)\\
\hline
    \end{tabular}
    \caption{length of video descriptions (in characters)}
    \label{tab:length}
\end{table}

\begin{figure}
  \includegraphics[width=0.45\textwidth]{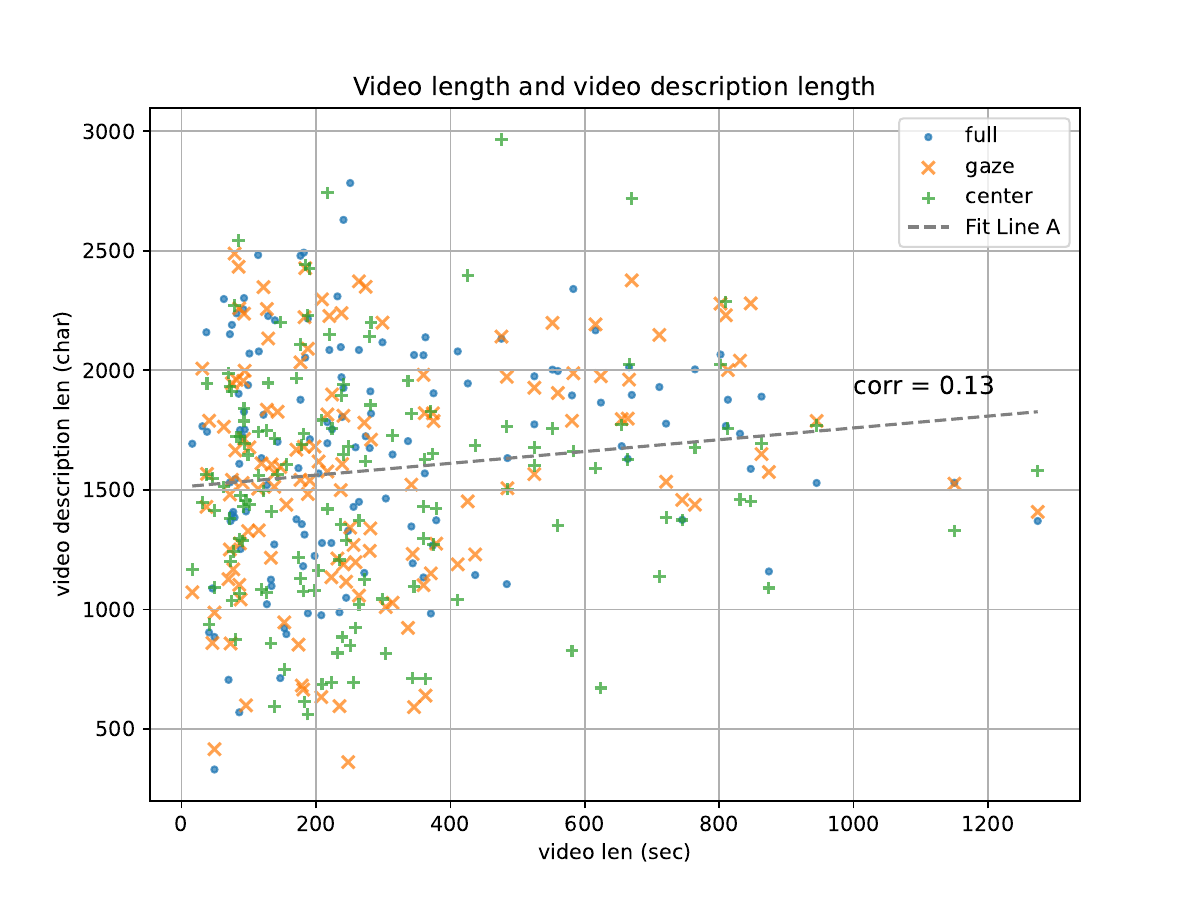}
  \caption{Video length and video description length}
  \label{fig:splat}
\end{figure}

\begin{figure*}
\centering
  \includegraphics[width=0.8\textwidth]{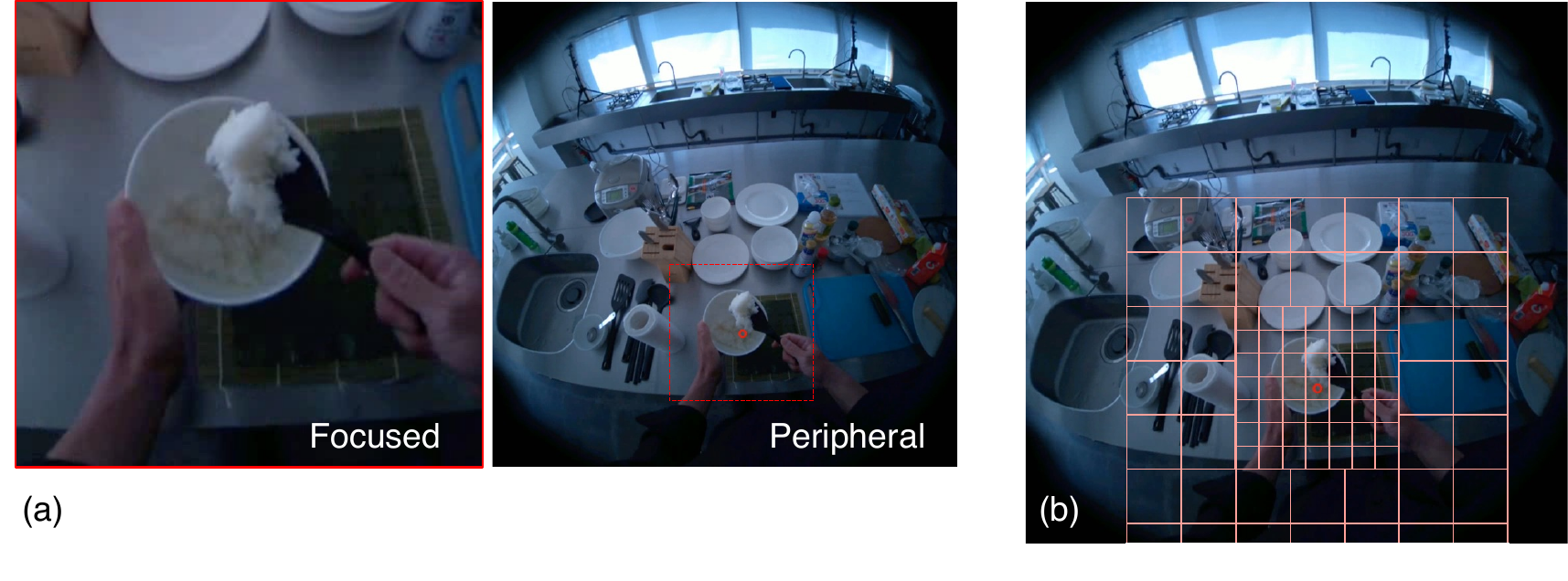}
  \caption{Alternative video encoding schemes for GazeLLM: (a) use focused and peripheral videos (b) adaptive ViT grid size}
  \label{fig:enc}
\end{figure*}

\begin{figure*}
\centering
  \includegraphics[width=0.8\textwidth]{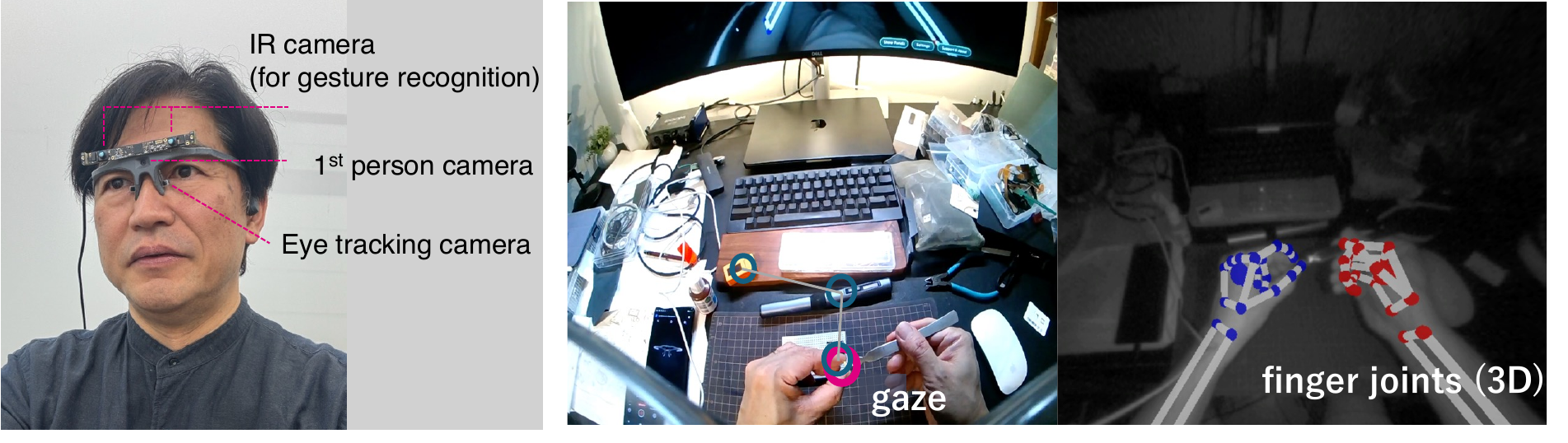}
  \caption{GazeLLM wearable configuration with 1st vision camera, eye-tracker, and hand gesture recognizer}
  \label{fig:gesture}
\end{figure*}

We investigated the relationship between video length and description length. No significant trends were observed in the character count of generated video descriptions across conditions (Full, Gaze, Center) (Table~\ref{tab:length}). In addition, there was only a weak correlation between video length and the character count of the descriptions (correlation coefficient: 0.13) as shown in Fig~\ref{fig:splat}. This suggests that, in the Full condition, the broader capture range compared to the Gaze and Center conditions includes more objects—many of which may not be directly related to the task—leading to potential declines in the quality of the descriptions as the model attempts to account for these additional elements.

Overall, the evaluation experiments confirmed the following:
\begin{itemize}
\item User assessments indicated that descriptions generated from videos under the Gaze condition received higher ratings compared to those from full 1st-person videos (Full condition) or videos cropped from the center of the full image (Center condition). Given that the pixel count in Gaze videos is only one-tenth of that in Full videos, this demonstrates effective information reduction as input for the MLLM.
\item In the four quantitative evaluations (BLEU, ROUGE, Sentence-BERT, and LLM-based evaluation), the Gaze condition also achieved higher scores than the Center condition, suggesting that gaze-based cropping is more effective.
\end{itemize}

\section{Discussions}

This study proposes a method to reduce data volume from 1st-person video by leveraging gaze information, allowing efficient comprehension by the LLM. Currently, we simply crop a rectangular region centered on the gaze point, but there is room for further improvement (Fig~\ref{fig:enc}). Fig~\ref{fig:enc} (a) illustrates an approach that provides both gaze-centered video and a downsampled full view, enabling comprehension of off-gaze areas at a lower resolution. This approach corresponds to the relationship between the central and peripheral fields in human vision. In MLLM research, methods such as Multimodal Rotary Position Embedding (M-RoPE)~\cite{Qwen2VL}, which provide unified positional encoding for multi-stream inputs with multiple temporal layers, are being explored. Applying M-RoPE to GazeLLM may be feasible.

Fig~\ref{fig:enc} (b) also illustrates another approach introducing a deformed grid for the 2D patches in ViT, where patches are denser near the gaze point and sparser toward the periphery. This also mirrors the relationship between the central and peripheral fields of human vision.

It is also possible to incorporate information about the wearer beyond gaze data. For example, objects being manipulated by the hands are likely relevant to the task, so hand position information could be prioritized for recognition by the LLM, similar to gaze information (Fig~\ref{fig:gesture}). Since gaze tends to anticipate the next point of action, combining hand activity data with gaze input could enable a more accurate understanding of actions.

\section{Conclusion}

In this study, we conducted experiments on a multimodal-LLM (MLLM) to generate task descriptions from 1st-person vision videos. Specifically, we evaluated how well tasks were described when images were cropped based on gaze, even though the cropped images contained only one-tenth of the pixels of the original. Six types of videos—spanning cooking, bike repair, healthcare, sports, and more (a total of 135 videos)—were evaluated. The assessment utilized BLEU, ROUGE, sentence-BERT metrics, and an LLM-based evaluation of descriptions. In all cases, videos cropped according to gaze received significantly higher evaluation scores compared to videos cropped from the image center.

User evaluations (25 participants) also confirmed that descriptions generated from gaze-based cropped videos were rated significantly higher than those from the videos cropped from the center, and were equivalent to the full videos.

These findings indicate that using gaze-based cropping when generating descriptions of 1st-person videos with MLLMs is an effective approach. This method reduces the number of pixels processed, thereby contributing to reduced computational load and memory usage for the LLM. This approach also suggests the potential for gaze-informed MLLMs to handle longer task videos efficiently.

\begin{acks}
 This work was supported by This work was supported by JSPS KAKENHI Grant 23K248948, JST Moonshot R\&D Grant JPMJMS2012, and JPNP23025 commissioned by the New Energy and Industrial Technology Development Organization (NEDO).
\end{acks}
\bibliographystyle{ACM-Reference-Format}
\bibliography{reference}

\end{document}